\documentclass{jps-cp}
\usepackage{txfonts} 

\title{Spin-Triplet Superconductivity in UTe$_2$ and Ferromagnetic Superconductors}

\author{
Dai~Aoki$^{1,2}$\thanks{E-mail: aoki@imr.tohoku.ac.jp}, 
Ai~Nakamura$^1$,
Fuminori~Honda$^1$,
DeXin~Li$^1$,
Yoshiya~Homma$^1$
Yusei~Shimizu$^1$,
Yoshiki~J.~Sato$^1$,
Georg~Knebel$^2$,
Jean-Pascal~Brison$^2$,
Alexandre~Pourret$^2$,
Daniel~Braithwaite$^2$, 
Gerard~Lapertot$^2$,
Qun~Niu$^2$,
Michal Vali\v{s}ka$^2$,
Hisatomo~Harima$^3$, and
Jacques~Flouquet$^2$
}

\inst{%
$^1$IMR, Tohoku University, Oarai, Ibaraki, 311-1313, Japan\\
$^2$University Grenoble Alpes, CEA, IRIG-PHELIQS, F-38000 Grenoble, France\\
$^3$Graduate School of Science, Kobe University, Kobe 657-8501, Japan
}

\email{aoki@imr.tohoku.ac.jp}

\recdate{September 16, 2019}

\abst{The spin-triplet state is most likely realized in uranium ferromagnetic superconductors, UGe$_2$, URhGe, UCoGe. 
The microscopic coexistence of ferromagnetism and superconductivity means that the Cooper pair should be realized under the strong internal field due the ferromagnetism.
leading to the spin-triplet state with equal spin pairing.
The field-reinforced superconductivity, which is observed in all three materials when the ferromagnetic fluctuations are enhanced, is one of the strong evidences for the spin-triplet superconductivity. 
We present here the results of a newly discovered spin-triplet superconductor, UTe$_2$, 
and compare those with the results of ferromagnetic superconductors.
Although no magnetic order is found in UTe$_2$, there are similarities between UTe$_2$ and ferromagnetic superconductors.
For example, the huge upper critical field exceeding the Pauli limit and the field-reentrant superconductivity for $H\parallel b$-axis are observed in UTe$_2$, URhGe and UCoGe.
We also show the specific heat results on UTe$_2$ in different quality samples, focusing on the residual density of states in the superconducting phase.
}

\kword{ferromagnetism, superconductivity, metamagnetism, reentrant superconductivity, spin triplet, specific heat}

\begin{document}
\maketitle
The coexistence of ferromagnetism and superconductivity attracts much attention because unconventional superconductivity with spin-triplet state is realized.~\cite{Aok19,Aok12_JPSJ_review}
In general, ferromagnetism and superconductivity are antagonistic because the large internal field
due to the ferromagnetism easily destroys the superconducting Cooper pairs in conventional superconductors. 
Thus it is natural to consider that the spin-triplet superconductivity with equal spin-pairing is realized in 
ferromagnetic superconductors.
The microscopic coexistence of ferromagnetism and superconductivity is established only in uranium compounds so far, namely UGe$_2$~\cite{Sax00}, URhGe~\cite{Aok01} and UCoGe~\cite{Huy07}.
All of these materials have fairly small ordered moments ($1$--$0.05\,\mu_{\rm B}/{\rm U}$) in the ferromagnetic phase compared to that for the U free ion ($\sim 3\,\mu_{\rm B}$). 
Thus the $5f$ electrons for these compounds are believed to be itinerant in the first principle,
although the magnetic anisotropy is rather strong, indicating the Ising properties. 
The superconductivity occurs well below the Curie temperature, $T_{\rm Curie}$, in the ferromagnetic state.
One of the highlights in ferromagnetic superconductors is the field-reentrant or field-reinforced superconductivity. 
In URhGe, for example, the reentrant superconductivity appears when the field is applied along the hard-magnetization $b$-axis in the orthorhombic structure.~\cite{Lev05} 
While the transition temperature $T_{\rm sc}$ is $0.25\,{\rm K}$ at zero field, 
the reentrant superconducting phase has a maximum $T_{\rm sc}$ of $0.4\,{\rm K}$ at $H_{\rm R}\sim 12\,{\rm T}$,
indicating that the superconductivity is indeed reinforced under magnetic field.
The similar field-reinforced superconductivity is also observed in UCoGe.~\cite{Aok09_UCoGe}

Recently a new spin-triplet superconductor, namely UTe$_2$ was discovered~\cite{Ran19,Aok19_UTe2}. 
UTe$_2$ has the body-centered orthorhombic structure with the space group $Immm$ ({\#}71, $D_{2h}^{25}$).
The distance of the first nearest neighbor for U atom is $3.78\,{\rm \AA}$, which is larger than the so-called Hill limit ($\sim 3.5\,{\rm \AA}$).
Although no magnetic order was found down to $0.025\,{\rm K}$, 
UTe$_2$ is believed to be at the verge of ferromagnetic order. 
In fact, the ferromagnetic fluctuations were observed in $\mu$SR~\cite{Sun19} and NMR experiments~\cite{Tok19}.
By substituting Te with Se, the ferromagnetic order appears at $69$ and $33\,{\rm K}$ in UTe$_{0.72}$Se$_{1.28}$ and UTe$_{0.24}$Se$_{1.76}$, respectively,~\cite{Noe96} 
although the space group for these materials is $Pnma$, which is different from that in UTe$_2$.
The superconducting transition occurs at $1.6\,{\rm K}$ with the sharp and large specific heat jump. 
The large residual density of states nearly $50\,{\%}$ may suggest the possibility for the spontaneous spin-polarization and the ``partially-gapped'' superconductivity similar to the A$_1$ state with non-unitary state.
However, it should be stressed that the direct transition from the paramagnetic state to the non-unitary state at zero field is forbidden from the symmetry restriction in this orthorhombic system.~\cite{Min19_comment}
Thus, the hidden feature in the superconducting state is expected. No other transition in the superconducting state is not observed yet at least at zero field.
The pressure study is definitely important to solve this problem. 
One of the strongest support for the spin-triplet superconductivity in UTe$_2$ is the huge upper critical field, $H_{\rm c2}$. 
In all the field directions, $H_{\rm c2}$ extremely exceeds the Pauli limit ($\sim 3\,{\rm T}$) expected for the weak-coupling BCS theory. 
The values of $H_{\rm c2}$ at $0\,{\rm K}$ are $7$ and $11\,{\rm T}$ for $H \parallel a$ and $c$-axis, respectively.
For $H\parallel b$-axis, the spectacular field-reentrant superconductivity is observed.~\cite{Kne19,Ran19_HighField}
The transition temperature monotonously decreases with field down to $0.4\,{\rm K}$ at $16\,{\rm T}$,
then increases with field up to $0.9\,{\rm K}$ at $35\,{\rm T}$.
The first order metamagnetic transtion occurs at $H_{\rm m}=35\,{\rm T}$~\cite{Kna19,Miy19}, and the superconductivity
is abruptly collapsed above $H_{\rm m}$.
The metamagnetic transition at $H_{\rm m}$ is connected to the so-called $T_{\chi, \rm max}$ at low fields,
where the magnetic susceptibility shows a broad maximum for $H \parallel b$-axis.
The magnetic susceptibility shows the Curie-Weiss behavior at high temperatures.
At low temperatures, the anisotropic susceptibility is observed with the relation, $\chi_a > \chi_c > \chi_b$, which is consistent with the anisotropy of $H_{\rm c2}$.

In order to study more details on superconducting properties in UTe$_2$, we have grown single crystals of UTe$_2$ with different quality, and measured the specific heat at low temperatures.
We compare the ($H,T$) phase diagrams for $H\parallel b$-axis in UTe$_2$, URhGe and UCoGe.

\begin{figure}[tbh]
\begin{center}
\includegraphics[width=.7 \hsize,pagebox=cropbox,clip]{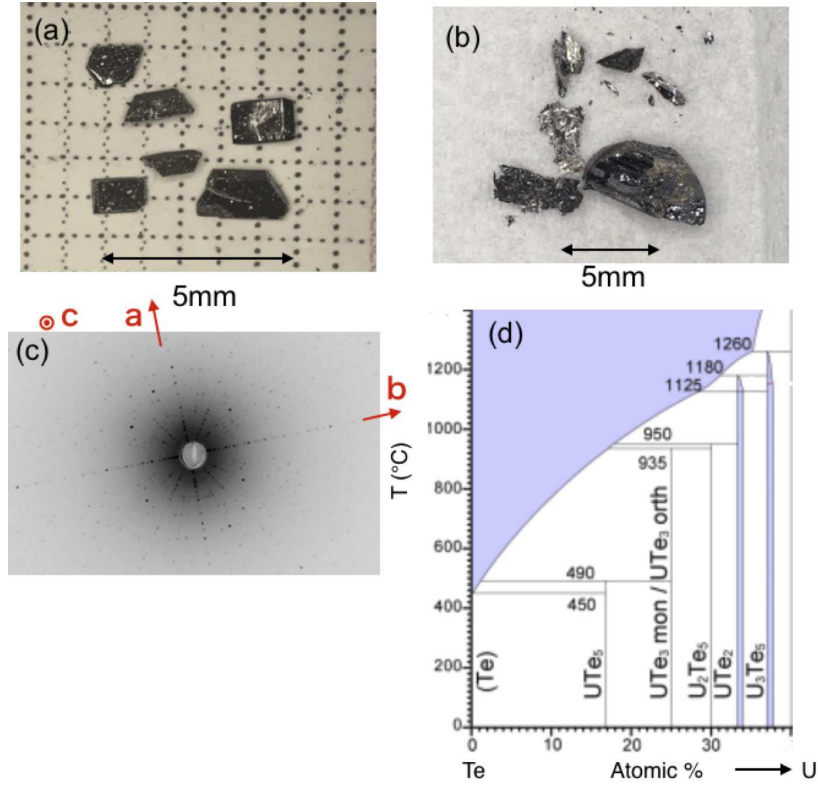}
\end{center}
\caption{(Color online) Photographs of UTe$_2$ single crystals grown by (a) chemical vapor transport method and (b) Te-flux method. (c) Laue photograph of UTe$_2$ single crystal along $c$-axis. (d) U-Te phase diagram cited from Ref.\citen{AtomWork}}
\label{fig:UTe2_photo}
\end{figure}
Single crystals of UTe$_2$ were grown using chemical vapor transport method.
The starting materials of U and Te were put into a quartz ampoule with the atomic ratio, U : Te = 2 : 3,
together with iodine as the transport agent to be the density, $3\,{\rm mg/cm}^3$ in the inner volume of the quartz ampoule. 
The ampoule was slowly heated and was kept at the temperature gradient of $1060\,^\circ {\rm C}$/$1000\,^\circ {\rm C}$ for 10 days.
Many single crystals were obtained at lower temperature side, as shown in Fig.~\ref{fig:UTe2_photo}(a).
The obtained single crystals were checked by the single crystal X-ray analysis.
The lattice parameters and the atomic coordinates are in good agreement with the values in the previous report.~\cite{Ike06_UTe2}
The single crystals were oriented using the Laue photograph, as shown in Fig.~\ref{fig:UTe2_photo}(c).
The clear superconducting transition was observed in resistivity and specific heat.
The highest residual resistivity ratio (RRR) is about 40.
Note that the we also obtained single crystals from the previous recipe, that is, a stoichiometric amount of starting materials and lower temperature gradient $950\,^\circ {\rm C}$/$850\,^\circ {\rm C}$.
The single crystals were grown at high temperature side in this case. 
However, the quality of the single crystal is lower with the low RRR ($\sim2$--$3$),
and no superconductivity was observed down to $0.1\,{\rm K}$.

As shown in Fig.~\ref{fig:UTe2_photo}(d), UTe$_2$ is an incongruent melting compound in the U-Te phase diagram, and single crystals of UTe$_2$ can be grown using the Te-flux method as well.
The off-stoichiometric amounts of U and Te ($22$ and $78\,{\rm at \%}$, respectively) were put into an alumina crucible, which was sealed in a Ta-tube under Ar atmosphere gas. 
The Ta-tube was then sealed again in a quartz ampoule.
The quartz ampoule was slowly heated up to $1050\,^\circ{\rm C}$ and was cooled down to $960\,^\circ{\rm C}$.
The Te-flux was removed at $960\,^\circ{\rm C}$ in a centrifuge.
The obtained single crystals were large, as shown in Fig.~\ref{fig:UTe2_photo}(b).
However the residual resistivity ratio is not very large (${\rm RRR}\sim 3$).
Although the superconductivity was confirmed by the resistivity, it was not a bulk property as no anomaly was detected in the specific heat. 
Hereafter, we show the results of single crystals obtained by the chemical vapor transport method with off-stoichiometric amounts of starting materials and the high temperature gradient.

\begin{figure}[tbh]
\begin{center}
\includegraphics[width=1 \hsize,pagebox=artbox,clip]{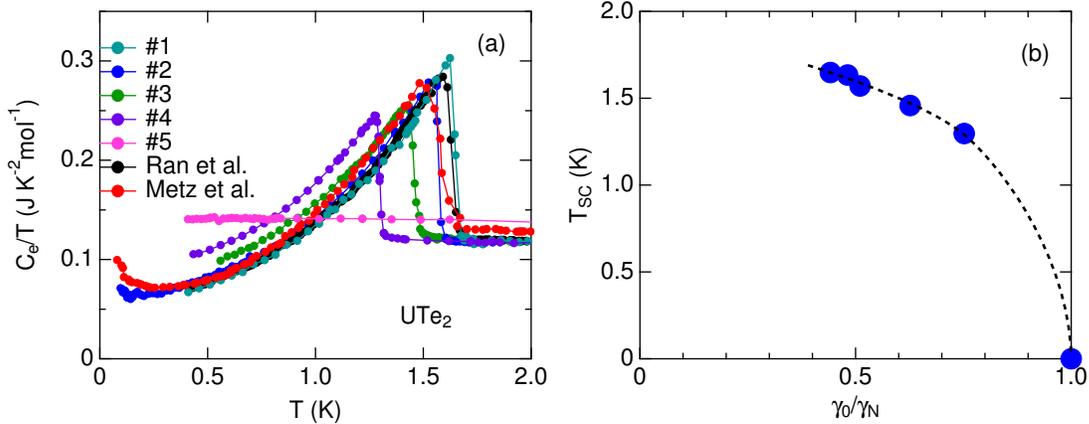}
\end{center}
\caption{(Color online) (a) Electronic specific heat in the form of $C_{\rm e}/T$ vs $T$ of UTe$_2$ in different samples. The phonon contribution is subtracted from the fitting at high temperature above $T_{\rm sc}$. Part of the data is replotted from Ref.~\citen{Ran19,Met19}.
(b) $T_{\rm sc}$ as a function of the normalized residual $\gamma$-value for different quality samples.}
\label{fig:UTe2_Cp}
\end{figure}

Figure~\ref{fig:UTe2_Cp}(a) shows the temperature dependence of the electronic specific heat in UTe$_2$. 
The part of the data is replotted from Ref.~\citen{Ran19,Met19}.
The data of sample {\#}1 show the highest $T_{\rm sc}$ with a sharp and large jump at $T_{\rm sc}$,
indicating the highest quality sample. 
The value of $T_{\rm sc}$ defined by the entropy balance in {\#1} is $1.65\,{\rm K}$,
and the residual $\gamma$-value, $\gamma_0$, which is extrapolated from the fitting $C_{\rm e}/T = \gamma_0 + \alpha T^2$ at low temperatures assuming a point node gap, is $\gamma_0=52\,{\rm mJ\,K^{-2} mol^{-1}}$. 
The residual $\gamma$-value is equal to $44\,{\%}$ of the $\gamma$-value in the normal state.
The lower quality samples show the lower $T_{\rm sc}$ and the higher residual $\gamma$-value.
For example, $T_{\rm sc}$ and $\gamma_0$ in sample {\#}4 are $1.23\,{\rm K}$ and $89\,{\rm mJ\,K^{-2}mol^{-1}}$, respectively.
In sample {\#}5, no superconductivity was observed down to $0.4\,{\rm K}$ in specific heat, and it is confirmed by the resistivity measurement down $0.1\,{\rm K}$.

Figure~\ref{fig:UTe2_Cp}(b) shows $T_{\rm sc}$ as a function of residual $\gamma$-value normalized by the $\gamma$-value in the normal state.
It is clear that $T_{\rm sc}$ decreases with increasing the residual $\gamma$-value. 
It is known that the decrease of $T_{\rm sc}$ can be described by the Abrikosov-Gor'kov pair-breaking theory.
On the basis of this model, the relation between $T_{\rm sc}$ and the residual density of states
had been studied theoretically~\cite{Oka11_Tsc} and experimentally~\cite{Kit94} in high $T_{\rm c}$ cuprates  and heavy fermion systems,
where the rapid increase of the residual density of states is reported, compared to the decrease of $T_{\rm sc}$.
This can be explained by the unitarity scattering in unconventional superconductivity.
The present result in Fig.~\ref{fig:UTe2_Cp}(b) supports the unconventional superconductivity in UTe$_2$.

An important question is whether the residual density of states exists in the ideal single crystal without impurities. 
In that case, the partial density of states would be gapped, and the so-called A1 state should be realized, where the time reversal symmetry must be broken at zero field.
There is, however, no experimental evidence for the breaking of time reversal symmetry in UTe$_2$.

\begin{figure}[tbh]
\begin{center}
\includegraphics[width=1 \hsize,pagebox=artbox,clip]{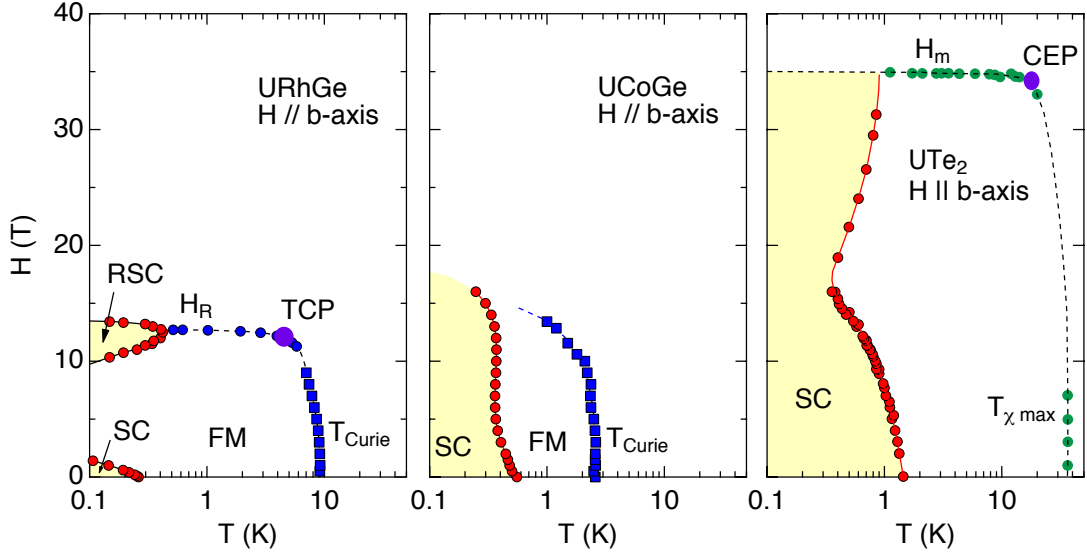}
\end{center}
\caption{(Color online) ($H,T$) phase diagrams for $H\parallel b$-axis in URhGe, UCoGe and UTe$_2$. The data are taken from Ref.~\citen{Aok19,Kne19,Kna19,Miy19}}
\label{fig:HT_phase}
\end{figure}

Next we show in Fig.~\ref{fig:HT_phase} the ($H,T$) phase diagrams of UTe$_2$ and two ferromagnetic superconductors, URhGe and UCoGe for $H\parallel b$-axis, corresponding to the hard-magnetization axis.
The field-reentrant or field-reinforced superconductivity is observed both in URhGe and in UCoGe.
The enhancement of superconductivity is clearly related to the suppression of $T_{\rm Curie}$,
where the ferromagnetic instabilities are realized.

In URhGe, the suppression of $T_{\rm Curie}$ leads to the spin-reorientation at $H_{\rm R}$ in field sweep at low temperatures.
The slope of magnetization curve for $H \parallel b$-axis is larger than that for $c$-axis.
The moment gradually tilts from $c$ to $b$-axis, and finally it re-orients to $b$-axis at $H_{\rm R}\sim 12\,{\rm T}$.
The $\gamma$-value is enhanced with increasing field, taking a maximum at $H_{\rm R}$.
In NMR experiments, the spin-spin relaxation rate, $1/T_2$ shows the diverging behavior around $H_{\rm R}$,
indicating the strong enhancement of the longitudinal ferromagnetic fluctuations~\cite{Tok15}.
The 2nd order transition of $T_{\rm Curie}$ at low fields changes into the weak 1st order transition at $H_{\rm R}$ through the tricritical point (TCP). 
The reentrant superconductivity appears with the maximum $T_{\rm sc}=0.4\,{\rm K}$ exactly at $H_{\rm R}$.

In UCoGe, the suppression of $T_{\rm Curie}$ with field is similar to the case for URhGe. 
However, the spin reorientation is not observed in magnetization curve, indicating the strong Ising property compared to URhGe.
The superconductivity shows an ``S''-shaped curve, which is also connected to the suppression of $T_{\rm Curie}$.
The enhancement of $\gamma$-value and the development of longitudinal fluctuation are observed in the field scan for $H\parallel b$-axis.

In UTe$_2$, the field reentrant superconductivity is also observed,
while the temperature range and field range are much wider, compared to those in ferromagnetic superconductors, URhGe and UCoGe.
The reentrant superconductivity is again linked to the metamagnetic transition at $H_{\rm m}$.
The clear difference from ferromagnetic superconductors is that
$H_{\rm m}$ at high fields originates from the broad maximum of magnetic susceptibility, $T_{\chi,\rm max}$, instead of $T_{\rm Curie}$. 
In the heavy fermion system, it is well known that $H_{\rm m}$ is scaled with $T_{\chi,\rm max}$~\cite{Aok13_CR}.
The value of $H_{\rm m}=35\,{\rm T}$ in UTe$_2$ is consistent with $T_{\chi,\rm max}=35\,{\rm K}$.
The mass enhancement around $H_{\rm m}$ is detected in the resistivity $A$ coefficient~\cite{Kna19} and
$\gamma$-value from the Maxwell relation in magnetization~\cite{Miy19} and the direct specific heat measurements~\cite{Ima19}.
The crossover at $T_{\chi,\rm max}$ changes into the 1st order transition at $H_{\rm m}$ through the critical end point (CEP).
It should be noted that the reentrant superconductivity is abruptly suppressed above $H_{\rm m}$ in UTe$_2$.
On the other hand, the reentrant superconductivity in URhGe still survives in the small field range above $H_{\rm m}$.
This is probably due to the abrupt change of $T_{\rm sc}$ as it is inferred from the sharp increase of magnetoresistance at $H_{\rm m}$,
implying the drastic change of the electronic state at $H_{\rm m}$.

In summary, we presented the single crystal growth of the novel spin-triplet superconductor, UTe$_2$,
and the results of specific heat in different quality samples. 
The higher quality sample shows the higher $T_{\rm sc}$ and the lower residual density of states. 
The rapid increase of the residual density of states compared to the decrease of $T_{\rm sc}$
supports the unconventional superconductivity in UTe$_2$.
The unusual field-reentrant superconductivity is a common feature in ferromagnetic superconductors and UTe$_2$. 
The precise high field experiments from the microscopic point of views and pressure experiments using a high quality single crystal are desired for the future studies.

\section*{Acknowledgements}
We thank Y.~Tokunaga, S.~Ikeda, Y. \={O}nuki, K. Ishida, K. Izawa, K. Miyake, V. Mineev, S. Ran, J. Ishizuka, Y. Yanase, K. Machida, C. Paulsen and K. Miyake
for fruitful discussion.
This work was supported by ERC starting grant (NewHeavyFermion), KAKENHI (JP15H05884, JP15H05882, JP15K21732, JP16H04006, JP15H05745, JP19H00646), and ICC-IMR.


\end{document}